\documentclass[oneside,final,11pt]{amsart}

\newcommand{\R}{\mathds R}
\newcommand{\dd}{\mathrm d}

\DeclareMathOperator{\Imag}{Im}

\usepackage{dsfont}

\title[Reply to ``Maximal violation of Bell inequalities'']{Reply to ``Maximal violation of Bell inequalities by position measurements''}

\author[D. Tausk]{Daniel V. Tausk}
\address{Departamento de Matem\'atica,\hfill\break\indent Universidade de S\~ao Paulo, Brazil}
\email{tausk@ime.usp.br} \urladdr{http://www.ime.usp.br/\~{}tausk}

\date{December 7th, 2010}

\begin{document}

\begin{abstract}
In a recent article \cite{KiukasWerner}, Kiukas and Werner claim to have shown that Bohmian Mechanics does not make the same empirical predictions
as ordinary Quantum Mechanics. More precisely, they have shown that ordinary Quantum Mechanics predicts maximal violations of the CHSH--Bell inequality
for a certain experiment in which only position measurements are performed on two noninteracting entangled free non-relativistic particles. Kiukas and
Werner claim that Bohmian Mechanics doesn't predict a violation of the CHSH--Bell inequality for that experiment. We explain that such claim is wrong
and that the argument supporting their claim neglects the fact that Bohmian Mechanics does not satisfy all the assumptions needed to prove the CHSH--Bell inequality.
We also clear up another few misconceptions about Bohmian Mechanics appearing in \cite{KiukasWerner}.
\end{abstract}

\maketitle

The CHSH inequality\footnote{%
After Clauser, Horne, Shimony and Holt \cite{CHSH}. The inequality is also sometimes referred to as ``Bell's inequality'', even though it is not the inequality
appearing in Bell's celebrated paper \cite{BellEPR}. Nevertheless, Bell himself used the CHSH inequality to explain his nonlocality argument in several papers (see
\cite{BellSpeakable}).} concerns an experiment in which, after a preparation procedure, two experimenters (normally referred to as Alice and Bob) can each
choose between two options for running an experiment having two possible outcomes, to which one assign the numbers $-1$ and $+1$.
The experiments are assumed to be performed at spacelike separation. Denoting Alice's outcome by $A\in\{-1,1\}$ and Bob's outcome by $B\in\{-1,1\}$, then the
CHSH inequality reads:
\[|E_{11}(AB)+E_{12}(AB)+E_{21}(AB)-E_{22}(AB)|\le2,\]
where $E_{ab}$ denotes expected value and $a,b\in\{1,2\}$ refer respectively to Alice and Bob's possible experimental choices. The CHSH inequality
can either be proved from {\em Bell's locality condition\/}\footnote{%
See \cite{Bellbeable, Bellsocks} and also \cite{Bellcuisine, Norsen} for further discussion on Bell's locality condition.} or from the assumption of the
existence of so called {\em non-contextual hidden variables}. In the present context, the latter assumption means that $A$ and $B$ can be written as functions
$A=A(a,\lambda)$, $B=B(b,\lambda)$ where, as above, $a,b\in\{1,2\}$ denote the possible experimental choices, and $\lambda\in\Lambda$ denotes a random
parameter belonging to a probability space $\Lambda$. {\em It is crucial for the proof of the CHSH inequality that $A$ not be allowed to depend on $b$ and $B$
not be allowed to depend on $a$}. It is precisely the fact that Bohmian Mechanics does not satisfy such assumption that Kiukas and Werner have overlooked.
If Alice's (resp., Bob's) experimental choices are associated to two $\{-1,1\}$-valued quantum observables represented by self-adjoint operators on a Hilbert
space $\mathcal H_A$ (resp., $\mathcal H_B$) having spectrum contained in $\{-1,1\}$ and the preparation procedure is associated to a quantum state on the
tensor product $\mathcal H_A\otimes\mathcal H_B$ then Quantum Theory predicts that the CHSH correlation (the lefthand side of the CHSH inequality) is bounded
by $2\sqrt2$. In their elegant paper \cite{KiukasWerner}, Kiukas and Werner present efficient techniques for determining conditions under which
there exist states that maximally violate the
CHSH inequality (i.e., for which the CHSH correlation equals $2\sqrt2$) as well as for explicitly determining such states.
Such techniques are then applied to obtain quantum states that maximally violate the CHSH
inequality for two noninteracting entangled free non-relativistic particles, where the quantum observables corresponding to the experimenters' choices are
$\{-1,1\}$-valued functions of the Heisenberg picture position operator at two distinct times. In other words, one obtains a maximal violation of the CHSH
inequality by position measurements alone\footnote{%
The hardest part of the work is the proof of the existence of {\em maximally\/} violating states. It is easy to prove the existence
of states violating CHSH from the assumption that Alice's and Bob's observables do not commute (see formula (3) of \cite{KiukasWerner}).}.

In Bohmian Mechanics, particles have trajectories $t\mapsto Q_k(t)\in\R^3$ which satisfy the first order ordinary differential equation:
\begin{equation}\label{eq:Bohm}
\frac{\dd Q_k}{\dd t}(t)=\frac\hbar{m_k}\Imag\frac{\psi^*\nabla_k\psi}{\psi^*\psi}\big(t,Q_1(t),\ldots,Q_n(t)\big),\quad k=1,\ldots,n
\end{equation}
where $\psi$ denotes the wave function satisfying the standard Schr\"odinger equation and $m_k$ the mass of the $k$-th particle.
Let $Q_A$ and $Q_B$ denote, respectively, the Bohmian trajectories of Alice's and of Bob's particle. Let us assume that, after the preparation of the initial
wave function, both particles are free (i.e., there is no potential term in Schr\"odinger's equation and no interaction with a measurement apparatus).
If one defines Alice's (resp., Bob's) outcome $A$ (resp., $B$) as being the appropriate $\{-1,1\}$-valued function of the Bohmian position $Q_A(t_a)$
(resp., $Q_B(t_b)$) at the appropriate instant $t_a$ (resp., $t_b$) that depends on Alice's (resp., Bob's) experimental choice $a\in\{1,2\}$ (resp., $b\in\{1,2\}$)
then the CHSH inequality {\em is not\/} violated: namely, if one takes the random parameter $\lambda$ to be the Bohmian positions $Q_A$, $Q_B$ at an instant
right after the preparation of the initial wave function then $A$ is indeed a function of $a$ and $\lambda$ (similarly, $B$ is a function of $b$ and $\lambda$),
so that all the assumptions needed to prove the CHSH inequality are satisfied. However, such observation is completely irrelevant for the issue of comparing the
empirical predictions of Bohmian Mechanics with those of ordinary Quantum Mechanics: namely, ordinary Quantum Mechanics has nothing to say about what happens to particles
when they are left unobserved. Ordinary Quantum Mechanics makes predictions for what Alice and Bob will see if they use a particle detector and {\em that\/} is the
experiment that one must analyze under Bohmian Mechanics in order to make a comparison. It turns out that, once the interaction of the particles with the detectors
are taken into account, Bohmian Mechanics does not satisfy the assumptions needed to prove the CHSH inequality as the interaction of a particle with a detector
influences not only the Bohmian trajectory of that particle but also the Bohmian trajectory of the distant particle (assuming that the particles are entangled),
so that now Bob's outcome $B$ is a function of $\lambda$ and of {\em both\/} Alice's choice $a$ and Bob's choice $b$\footnote{%
Since, according to Bohmian Mechanics, Bob's outcome $B$ depends on Alice's choice $a$, one might
get the impression that Bohmian Mechanics predicts that Alice can send superluminal messages to Bob. However, just like in ordinary Quantum Mechanics, such superluminal
messaging is not possible. In order to ``read'' Alice's message (the $a$), Bob would have to know more about the value of $\lambda$ --- the initial
Bohmian configuration --- than it is possible for him
to know without disturbing the preparation of the initial wave function (see \cite[Sections 4.2, 4.3]{BohmInfo} for a succinct exposition and \cite{QE} for a more
detailed one).}. It turns out that Bohmian Mechanics makes the same experimental predictions as ordinary Quantum Mechanics for that experiment\footnote{%
That this is indeed the case follows, for instance, from the following observation: if $t$ is an instant after which both experiments have been performed then
the Bohmian particle configuration of the apparatuses is $\vert\psi(t)\vert^2$-distributed and such configuration records the outcomes of both experiments.}
and that Bohmian Mechanics predicts a (maximal) violation of CHSH when ordinary Quantum Mechanics does so.

While the formulation of Bohmian Mechanics does not mention any ``wave function collapse'' and the wave function of a Bohmian universe evolves exclusively
through the Schr\"odinger equation, it turns out that, as a mathematical consequence of the dynamical equations of the theory,
the wave function of a {\em subsystem\/} of a Bohmian universe\footnote{%
The wave function of a subsystem is obtained by plugging into the wave function $\psi$ of the Bohmian universe the actual Bohmian positions of the particles not belonging to
that subsystem. Some technical complications arise in the case of particles with spin which we can safely ignore for the purpose at hand.}
{\em does not\/} always evolve through the Schr\"odinger equation, but it sometimes goes through the standard textbook ``collapse under measurement''
(see \cite{Dialogue, BohmInfo} for a succinct exposition and \cite{QE} for a more detailed one). According to Kiukas and Werner (\cite[Section II]{KiukasWerner}):
\begin{quotation}
``The simplest position [the proponents of Bohmian Mechanics can adopt in order to make the agreement with ordinary Quantum Mechanics possible]
is to include the collapse of the wave function into the theory. Then the first measurement instantaneously collapses the wave function.
So if agreement with quantum mechanics is to be kept, the probability distribution changes suddenly. There is no way to fit this
with continuous trajectories: When the guiding field collapses, the particles must jump.''
\end{quotation}
So, contrary to Kiukas and Werner's claim, one does not have to add any collapse rule into Bohmian Mechanics, for, as explained above,
the collapse of the wave function of a subsystem follows from the standard dynamical equations of the theory. Moreover, contrary to their claim,
there is no problem in fitting such collapse with continuous trajectories, as Bohmian trajectories are smooth. One should point out that
the wave function of a subsystem of a Bohmian universe does not go under any {\em instantaneous\/} collapse: namely, the wave function of a subsystem
is a smooth function of both time and configuration point and collapse happens over a positive amount of time.

Again, according to Kiukas and Werner (\cite[Section II]{KiukasWerner}):
\begin{quotation}
``This may be the reasons why many Bohmians adopt a\break strongly contextual view. In this view
one has to describe the measurement devices explicitly in the same theory, so all trajectories
depend on the entire experimental arrangement. Therefore, the trajectory probabilities in two
experiments, in which the measurements on particle A happen at different times, have no relation
to each other, not even for trajectories of particle B. So the two-time correlations computed from
the two-particle ensemble of trajectories are never observed anyhow, and hence pose no threat to
the theory. The downside of this argument is that it also applies to single time measurements, i.e.,
the agreement between Bohm--Nelson configurational probabilities and quantum ones is equally
irrelevant. The naive version of Bohmian theory holds ``position'' to be special, even ``real,'' while
all other measurement outcomes can only be described indirectly by including the measurement
devices. Saving the Nelson--Bohm theory's failure regarding two-time two-particle correlations by
going contextual also for position just means that the particle positions are declared {\em unobservable
according to the theory itself}, hence truly hidden.'' [emphasis in the original]
\end{quotation}
Definitely, the particle positions are not at all hidden according to Bohmian Mechanics. Indeed, according to the theory, ordinary matter is made precisely
out of the Bohmian particles, so that the configuration of ordinary macroscopic objects --- such as pointers, tables and chairs --- is precisely a Bohmian
configuration (and the fact that such configurations are equally distributed according to both Bohmian and ordinary Quantum Mechanics does imply
the empirical equivalence between the two theories\footnote{%
Keep in mind also that, for macroscopic objects, the Bohmian behavior approximates the classical one so that, for instance, one can neglect the effects on a table
caused by position measurements such as those done by an individual that looks at the table.}).
Moreover, in a Bohmian universe, ordinary particle detectors do find Bohmian particles precisely where they are,
i.e., particle detection is a genuine measurement of the Bohmian particle position. Of course, in order to compute the Bohmian trajectory of a particle, one has
to take into account everything that might influence that trajectory. This, of course, is not a feature of Bohmian Mechanics, but simply the right way of making
predictions using a physical theory: if, for instance, in Classical Mechanics, one computes the trajectory of a golf ball and do not take into account
that it banged against a tree, then one won't make the right predictions concerning the trajectory of that golf ball if it really did bang against
the tree. If taking into account all the relevant influences is what is meant by a ``strongly contextual view'' then it seems to be a pretty reasonable view.
The relevant difference between Bohmian Mechanics and Classical Mechanics in this respect is that, while in Classical Mechanics one can arrange position
measurements whose effect on the trajectory can be made arbitrarily small, the same is not true for Bohmian Mechanics. Any experiment that works
as a position measurement will collapse the wave function of the particle whose position is being measured and thus significantly disturb its trajectory.
Thus, if $t\mapsto Q(t)$ denotes the Bohmian trajectory of a non-observed particle, then a position measurement at time $t$ yields $Q(t)$. On the other
hand, if one performs {\em both\/} a position measurement at time $t_1$ and a position measurement at time $t_2>t_1$ then the first measurement yields
$Q(t_1)$ and the second yields not $Q(t_2)$ but $\widetilde Q(t_2)$, where $\widetilde Q$ is the trajectory computed by taking into account the measurement
at time $t_1$. At time $t_2$ one finds the particle where {\em it really is}, not (obviously)
where it would have been had the first measurement not been performed. Moreover, the fact that interaction with experimental equipment significantly
disturbs the behavior of the observed system is normally taken to be one of the main lessons of Quantum Theory and it is empirically confirmed by,
say, the double slit experiment, in which placing a detector on one of the slits dramatically changes the detection pattern on the screen.

Finally, one should observe that, according to Bohmian Mechanics, if two particles are entangled then what happens to one particle influences
the other particle. Indeed, observe that, according to equation \eqref{eq:Bohm}, the velocity of the $k$-th particle might depend on the position of all the other
particles\footnote{%
However, it also follows easily from equation \eqref{eq:Bohm} that if two particles are not entangled then the differential equation that defines the Bohmian
trajectory of one particle does not involve the trajectory of the other particle.}.
Thus, in the kind of experiment considered by Kiukas and Werner, Alice's position measurement is relevant for computing the Bohmian trajectory of Alice's particle
and hence also for computing the Bohmian trajectory of Bob's particle. As one should expect, Bob will find his particle where it really is and not where it would have
been had Alice's measurement not been performed. This surprising ``spooky'' action at a distance predicted by Bohmian Mechanics is, by the way, also a fact
of nature: namely, as shown by Bell (see, for instance, \cite{Bellcuisine}), one cannot account for the violation of the CHSH inequality within a local theory.

\end{document}